\documentclass[twocolumn]{aastex631}

\usepackage{soul}
\usepackage{mathtools}
\usepackage{empheq}
\usepackage{comment}
\usepackage{bbm}
\usepackage{mathrsfs}
\usepackage{amsfonts}
\usepackage{color} 
\usepackage{pbox}
\usepackage{xcolor} 
\usepackage{comment}
\usepackage{tensor}
\usepackage{physics}
\usepackage{listings}
\usepackage{multirow}

\usepackage{hyperref}
\hypersetup{
    colorlinks=true,
    linkcolor=black,
    filecolor=magenta,      
    urlcolor=cyan,
    pdftitle={Overleaf Example},
    pdfpagemode=FullScreen,
}
\allowdisplaybreaks

\begin{document}

\title{An NGC 1068-Informed Understanding of Neutrino Emission of the Active Galactic Nucleus TXS~0506+056}

\author[0000-0002-8735-8579]{Arifa Khatee Zathul}
\affiliation{Department of Physics, University of Wisconsin--Madison, \\ Madison, Wisconsin, 53706, USA}

\author{Marjon Moulai}
\affiliation{Department of Physics, University of Wisconsin--Madison, \\ Madison, Wisconsin, 53706, USA}

\author[0000-0002-5387-8138]{K.~Fang}
\affiliation{Department of Physics, University of Wisconsin--Madison, \\ Madison, Wisconsin, 53706, USA}

\author[0000-0001-6224-2417]{Francis Halzen}
\affiliation{Department of Physics, University of Wisconsin--Madison, \\ Madison, Wisconsin, 53706, USA}

\date{\today}

\begin{abstract}
We present arguments that the neutrinos observed by IceCube from the active galactic nucleus TXS~0506+056 may originate near its core and not in the blazar jet. The origin of the neutrinos is consistent with the mechanism that produces the neutrino flux observed from the active galaxies NGC~1068 and NGC~4151, but requires an Eddington luminosity cosmic ray flux to compensate for its larger distance. Like NGC~1068, the source is characterized by episodes of high X-ray emission and is gamma-ray-obscured during the 2014 burst, and there is evidence that this is also the case during the short burst in 2017 that produced IC-170922. The observations may be explained as a flux originating in an obscured core within $10 \sim 100$ Schwarzschild radii from the central black hole, which is not transparent to gamma rays from neutral pions accompanying the neutrinos.
\end{abstract}


\section{Introduction}
\label{sec:intro}

Understanding the high-energy neutrino emission of the ``blazar" TXS~0506+056, the first cosmic ray accelerator identified in a multimessenger campaign in 2017, has presented an unmet challenge. Guided by its astronomical classification as a BL Lac object, modeling of neutrino production has been focused on the jet. Jets emitting high-energy photons are unlikely to be intense neutrino sources. Transparent to photons, protons, even if accelerated along with electrons, are unlikely to find a target to produce neutrinos since the number of protons available for interaction might be limited due to the jet's low opacity to protons.
Detailed modeling has confirmed that matching the neutrino flux level of the 2017 burst is challenging; the origin of the gamma-ray-obscured 2014 burst remains a mystery (see, for instance, \citealp{wang2024unifiedmodelmultiepochneutrino,Das_2022,Acciari_2022,Jin_2021,Sahu_2020,Cao_2020,Zhang_2020,winter2019multimessengerinterpretationneutrinostxs,Xue_2019,Padovani_2019,Reimer_2019,Wang_2022,Sahakyan_2018,Keivani_2018,Cerruti_2018}). While multi-zone models leveraging a large number of parameters may fit the data, we argue that the mechanism that naturally accommodates the production of neutrinos in the core of the active galaxy NGC~1068 may also contribute to the neutrino production in TXS~0506+056. 

A neutrino source requires the acceleration of protons and the presence of a neutrino-producing target. These two requirements may naturally occur close to the black holes at the centers of active galaxies, as underscored by the observation of NGC~1068~\citep{IceCube:2022der}. 
NGC~1068 is among the brightest nearby active galaxies. Time-integrated searches using IceCube data have identified an excess of high-energy neutrinos from its direction~\citep{icecube2022evidence}. The energy flux of neutrinos is one to two orders of magnitude higher than the GeV to TeV gamma-ray flux of this active galaxy. This requires a neutrino production site that is opaque to high- and very-high-energy gamma rays, narrowing the possibilities down to the core of the active galaxy~\citep{murase2020hidden, Inoue:2019yfs}.

Neutrino production by cosmic ray accelerators requires a high-density target of photons or protons to produce pions or other particles that are the parents of neutrinos, such as X-ray coronas or regions of large hydrogen density near the cores of active galaxies. Among active galaxies, intense X-ray emission and large optical depth, $N_H$ are likely to be better indicators of neutrino emission than their astronomical classification, which is based on their orientation relative to the observer and unlikely to be relevant.

We briefly summarize here the relevant multimessenger information on TXS~0506+056 before interpreting it in the broad context of the corona model for neutrino production in active galaxies~\citep{Das:2024vug, Mbarek_2024, Fiorillo_2024, fiorillo2024b, Murase:2022dog, Kheirandish:2021wkm, Eichmann:2022lxh, Inoue:2019yfs, Inoue:2021iyl, Ajello:2023hkh, Anchordoqui:2021vms} inspired by the observations of NGC~1068. On September 22, 2017, IceCube reported a well-reconstructed muon that deposited 180 TeV inside the detector, corresponding to a most probable energy of 290\,TeV~\citep{kopper2017icecube,icecube2018multimessenger} for the parent neutrino, known as IC-170922A. Its arrival direction was aligned with the coordinates of a known {\it Fermi} blazar, TXS~0506+056, to within $0.06^\circ$. The source was ``flaring" with a gamma-ray flux that had increased by a factor of seven in recent months. A variety of estimates converged on a probability on the order of $10^{-3}$ that the coincidence was accidental~\citep{kopper2017icecube, icecube2018multimessenger}. The identification of the neutrino with the source reached the level of evidence, but not more. Searching the archival IceCube data revealed that the neutrino luminosity of TXS~0506+056 is dominated by a burst observed in 2014-2015~\citep{icecube2018neutrino} which leaves the burst associated with IC170922 as a subdominant contribution. What clinched the association was a series of subsequent observations, culminating with the optical observation of the source switching from an ``off" to an ``on" state two hours after the emission of IC-170922A~\citep{lipunov2021vizier}, conclusively associating the neutrino with TXS~0506+056 in the time domain. 

The sequence of observations relevant to this paper can be summarized as follows:
\begin{itemize}
    \item The redshift of the host galaxy was measured to be $z\simeq0.34$~\citep{paiano2018redshift}. It is important to realize that nearby blazars like the Markarian sources are at a distance that is ten times closer, and therefore TXS~0506+056, with a similar gamma ray flux despite its greater distance, is one of the most luminous sources in the universe. This suggests that it belongs to a special class of sources that accelerate proton beams inside dense environments, as revealed by the neutrino.
    \item Originally detected by NASA's {\it Swift}~\citep{evans2017further} and {\it Fermi}~\citep{tanaka2017fermi} satellites, the neutrino alert was followed up by ground-based air Cherenkov telescopes~\citep{mirzoyan2017first}. MAGIC detected the emission of gamma rays with energies exceeding 100 GeV starting several days after the observation of IC-170922A~\citep{ansoldi2018blazar}. Given its distance, this establishes the source as a relatively rare TeV blazar.
    \item Informed on where to look, IceCube searched its archival neutrino data up to and including October 2017 for evidence of neutrino emission at the location of TXS~0506+056. IceCube found 19 high-energy neutrino events in 2014-2015 
    on a background of fewer than six in a burst that lasted 110 days~\citep{icecube2018neutrino}.
    This burst dominates the integrated flux from the source over the last 9.5 years of archival IceCube data, leaving the 2017 flare as a second subdominant feature.

    \item The MASTER robotic optical telescope network, which had been monitoring the source since 2005, found its strongest time variation in the last 15 years to have occurred two hours after the emission of IC-170922, with a second less statistically compelling variation following the 2014 burst~\citep{lipunov2020optical}. The source switches from the ``off" to the ``on" state two hours after the emission of the neutrino. After monitoring the uniformity of their observations until the first quarter of 2020, they concluded that the time variation detected on September 22, 2017, conclusively associates the source with the neutrino at a level of $50\, \sigma$. The optical flash observed two hours after the 290-TeV neutrino not only established their association in the time domain but may indicate some rearrangement in the core or the accretion disk, hinting at a neutrino origin in the core rather than the jet of the galaxy.

    \item Using ten years of IceCube data collected between April 6, 2008 and July 10, 2018, a search for pointlike sources found TXS~0506+056 as the second most significant source, after NGC~1068, in IceCube's northern source catalog \citep{IceCube:2019cia}. The signal in the direction of TXS~0506+056 integrated over ten years has a pre-trial significance of $3.6\,\sigma$.

\end{itemize}

Additionally, it is important to note that the high-energy neutrino spectra covering the 2014 burst as well as the ten-year time-integrated search are consistent with a hard $\sim E^{-2}$ spectrum, which is expected for a cosmic accelerator.

In what follows, the motivation for a common origin of high-energy neutrinos from NGC~1068 and TXS~0506+056 will be mostly based on their high X-ray fluxes, but also that both sources are gamma-ray-obscured. This should not come as a surprise because it is also the case for other sources that have reached the level of evidence in the IceCube data~\citep{abbasi2024icecubesearchneutrinoemission,abbasi2024searchneutrinoemissionhard}. The high level of the diffuse neutrino flux observed by IceCube relative to the diffuse gamma-ray flux measured by the {\it Fermi} satellite indicates that typical neutrino sources are gamma-ray-obscured sources~\citep{Murase:2015xka,Fang_2022}. Where TXS~0506+056 is concerned, the dominant flare observed in 2014 is not accompanied by an elevated gamma-ray flux. What about the ``flaring" blazar in 2017? Although the source had been flaring for several months, the {\it Fermi} data indicates a minimum of the flux at the time IC-170922 was emitted; for a discussion, see \citet{Kun_2021}, with a hint that the source may have been gamma-ray-obscured over a short period, possibly hours, as suggested by the optical observation. In this paper, we will argue that the neutrinos are produced in the core of TXS~0506+056, which provides an appropriate environment for accelerating protons and converting them to neutrinos, as is the case in NGC~1068. 
 
Like many other blazars, the spectral energy distribution of TXS~0506+056 presents two emission peaks. The lower one peaking in the optical band extending to soft X-ray may be explained as synchrotron radiation by relativistic electrons from the jet. The second peak at gamma-ray energies is usually interpreted as the synchrotron-self-Compton emission by the same jet electrons or, in a hadronic scenario, as the electromagnetic cascades from photopion or Bethe-Heitler interactions of cosmic-ray protons from the jet (see example \citealp{keivani2018multimessenger}). On the other hand, recent  observations of blazar outbursts have identified an additional component in the X-ray spectrum extending well above 10 keV, which could represent a temporarily enhanced emission from an accretion disk corona~\citep{Komossa:2021exd}. Moreover, studies have shown that the X-ray emission in many radio-loud AGN appears to come from a corona, rather than from the jet  \citep{10.1093/mnras/staa1411}. The coronal X-ray emission from AGN is highly variable with flares on timescale of hours, reaching 100~keV and higher energies \citep{2018A&A...614A..37T}.

Hard X-ray emission from TXS~0506+056 was detected by NuSTAR in two Target of Opportunity observations following IC-170922, but unknown during the 2014 flare. Inspired by the blazar outburst observations, we argue that the corona of TXS~0506+056 may also partly contribute to the X-ray emission, which is at the level of $L_X\sim 9\times 10^{44}\,\rm erg\,s^{-1}$ at $\sim 15-55$~keV \citep{keivani2018multimessenger,kun2024correlation}. Importantly, the coronal power may exceed the source luminosity due to gas absorption, in particular observed in the soft X-ray band~\citep{Hickox:2018xjf}.

We will show that the neutrino flux observed from TXS~0506+056 can be accommodated assuming a core with characteristic photon densities that are similar to those in NGC~1068. The source does require a proton flux increased to the Eddington level during the neutrino flaring time in order to accommodate a neutrino flux similar to the one of NGC~1068 for a source at a larger distance.

\section{Corona-disk model}

A corona is a dense X-ray emitting region which surrounds the black hole. Assuming a spherical geometry with  a radius $R$, the energy density of the X-rays associated with the corona, $u_{X}$, is related to the X-ray luminosity, $L_X$, of the source by
\begin{align}
    u_{X}
    =
    \frac{ L_{X}
    }{
    4\pi c R^2
    } 
    \approx
    n_{X} E_{X}\,,
    \label{eq: energydensity}
\end{align}
where $L_X$ is the X-ray luminosity measured at energy $E_X$, $c$ is the speed of light, and $n_{X} \approx u_X/ E_X$ is the number density of X-rays in the target. Protons accelerated near the black hole interact with the X-ray photons to produce neutrinos. 

Protons may be accelerated in turbulence or collisionless magnetic reconnection in the coronal region. The coronal magnetic field, $B$, may be estimated by $u_B = B^2/8\pi = \xi_B u_X$, where $u_B$ is the magnetic energy density and $\xi_B$ is the conversion factor between radiation and magnetic energy density that is on the order of unity. The maximum attainable energy for protons is $E \sim eBR = e (2 L_X / \xi_B c)^{1/2}$, sufficient to produce the TeV neutrinos. The actual maximum proton energy and the spectrum depend on the acceleration mechanism. In a scenario where protons obtain their energies by stochastic scattering on the turbulent fluctuations, particles may be efficiently confined by the magnetic fields before cooling \citep{murase2020hidden,Mbarek:2023yeq,2024arXiv240701678F}. In a scenario where protons are accelerated by magnetic reconnection, the short acceleration process competes with a fast escape of the particles from the reconnection layer \citep{Fiorillo:2023dts}. A two-step acceleration process where particles get pre-accelerated before arriving at the corona may also be present \citep{Mbarek:2023yeq}.  

Relativistic protons leave the acceleration region by diffusing away from the turbulent magnetic field or being advected towards the black hole. 
The fallback time is $t_{\rm fall} = R / \alpha v_K$, where $\alpha \sim 0.1$ is the viscous parameter and $v_K = (GM/R)^{1/2}$ is the disk velocity in circular Keplerian orbits \citep{shakura1973black}. The diffusion time, $t_{\rm diff}$, depends on the strength and properties of the magnetic field. In general, $t_{\rm fall}$ is shorter than $t_{\rm diff}$ at lower energies and longer than $t_{\rm diff}$ at higher energies. As in  \citet{Stathopoulos:2023qoy, Das:2024bed}, we assume that the escape time of charged particles is a constant factor of the light crossing time, $t_{\rm esc} = (c/v_{\rm esc}) \, (R/c)$, where the ratio between the light speed and the escape speed of charged particles is $c/v_{\rm esc}= 10-100$.

The critical quantity for neutrino production is the proton's opacity for interacting in the corona. It determines the number of pions that decay into neutrinos and gamma rays:

\begin{equation}\label{eqn:tau_pg}
    \tau_{p\gamma} = \kappa_{p\gamma} \frac{c t_{\rm esc}}{\lambda_{p\gamma} }  \approx n_X  \kappa_{p\gamma} \sigma_{p\gamma} c t_{\rm esc} \approx \frac{u_X}{E_X} \kappa_{p\gamma} \sigma_{p\gamma} c t_{\rm esc} \,.
\end{equation}

We will assume that the pions are produced on the $\Delta$ resonance with inelasticity $\kappa_{p\gamma} \equiv E_{\pi}/E_p \simeq 0.2$, which encodes the amount of energy transferred from protons to pions ($\pi^0$ or $\pi^+$) in each interaction. $\lambda_{p\gamma}$ is the interaction length, which is inversely proportional to the density of the target,  $n_{X}$, and to the photoproduction cross section, $\sigma_{p\gamma} \simeq 5 \times 10^{-28}
\text{ cm$^2$}$. The optical depth can be rewritten in terms of the Eddington luminosity, $L_{\rm edd}$, and the Schwarzschild radius, $R_S$, of the black hole of mass $M$:
\begin{eqnarray}
    \tau_{p\gamma}
    &=&
    \frac{\kappa_{p\gamma} \sigma_{p\gamma} }{4\pi c}
    \Big(
    \frac{c}{v_{\rm esc}} 
    \Big) 
    \frac{1}{R}
    \frac{L_X}{E_X}
    \nonumber \\
    &\simeq&
    70 \,
    \Big(
    \frac{v_{\rm esc}}{c}
    \Big)^{-1}
    \Big(
    \frac{R}{R_s}
    \Big)^{-1}
    \Big(
    \frac{ E_{X} }{ 1 \text{ keV} }
    \Big)^{-1}
    \Big(
    \frac{ L_{X} }{ L_{\rm edd} }
    \Big)
    \,,
    \label{eq: tau}
\end{eqnarray}
where $L_{\rm edd} = 1.26 \times 10^{38} ( {M}/{ M_\odot})\, \rm erg\,s^{-1}$ and $R_s = 2GM/c^2 = 3\times 10^5\,M/M_\odot\,\rm cm$.

\begin{table*}[t!]
\begin{center}
\begin{tabular}{ |p{4.6cm}|p{4.6cm}|p{4.6cm}|  } 
\hline
Parameter & NGC~1068 & TXS~0506+056
\\
\hline
Mass, $M$ & 
$1.3^{+4.3}_{-0.6} \times 10^{7} M_\odot$ & 
$3.1^{+29.9}_{-2.7} \times 10^{8} M_\odot$  \\ 
\hline
Schwarzschild radius, $R_S$ & $3.9^{+5.1}_{-1.8} \times 10^{12}$ cm & 
$9.3^{+71.1}_{-8.2} \times 10^{12}$ cm  \\ 
\hline
Luminosity distance, $d_L$ & $14.4^{+2.1}_{-7.2}$ Mpc & $1774 ^{+63}_{-24}$ Mpc   \\ 
\hline
Eddington luminosity, $L_{\rm edd}$ & 
$1.6^{+2.1}_{-0.8} \times 10^{45}$ erg/s  & 
$3.9^{+29.9}_{-3.4} \times 10^{46}$ erg/s  \\ 
\hline
Neutrino luminosity, $L_\nu$ 
& $2.9^{+1.4}_{-2.4} \times 10^{42}$ erg/s 
& $6.8^{+4.4}_{-3.6} \times 10^{45}$ erg/s (9.5 year duration), $1.1^{+0.5}_{-0.4} \times 10^{47}$ erg/s (158 days duration) \\ 
\hline
X-ray luminosity, $L_X$ 
& $2.2^{+0.4}_{-1.7} \times 10^{43}$ erg/s 
& $8.5^{+2.0}_{-4.8} \times 10^{44}$ erg/s  \\ 
\hline
X-ray energy, $E_X$ 
& 1 keV
& 1 keV \\
\hline
\end{tabular}
\caption{Input values for modeling and calculating the opacities and the proton luminosities of NGC~1068 and TXS~0506+056.}
\label{table:taupg}
\end{center}
\end{table*}

In what follows we estimate the opacity using the X-ray flux at a typical energy of $1$ keV which lies within the soft X-ray energy band of 0.3 -- 10 keV.
The information on both sources that will be used as input into our modeling is summarized in Table~\ref{table:taupg}. Given the challenges in isolating the coronal X-ray flux contribution from the observed flux and the need to assign a neutrino flux to TXS~0506+056 based on only two observed neutrino flares, we have assigned generous uncertainties to the input quantities. Note that the uncertainties in the distance $d_L$ and Schwarzschild radius $R_s$ in Table~\ref{table:taupg} are not accounted for in the calculations.
The error propagation is done by separately summing the upper and lower errors in quadrature.

\begin{figure}[t]
\centering
\includegraphics[width=0.5\textwidth]{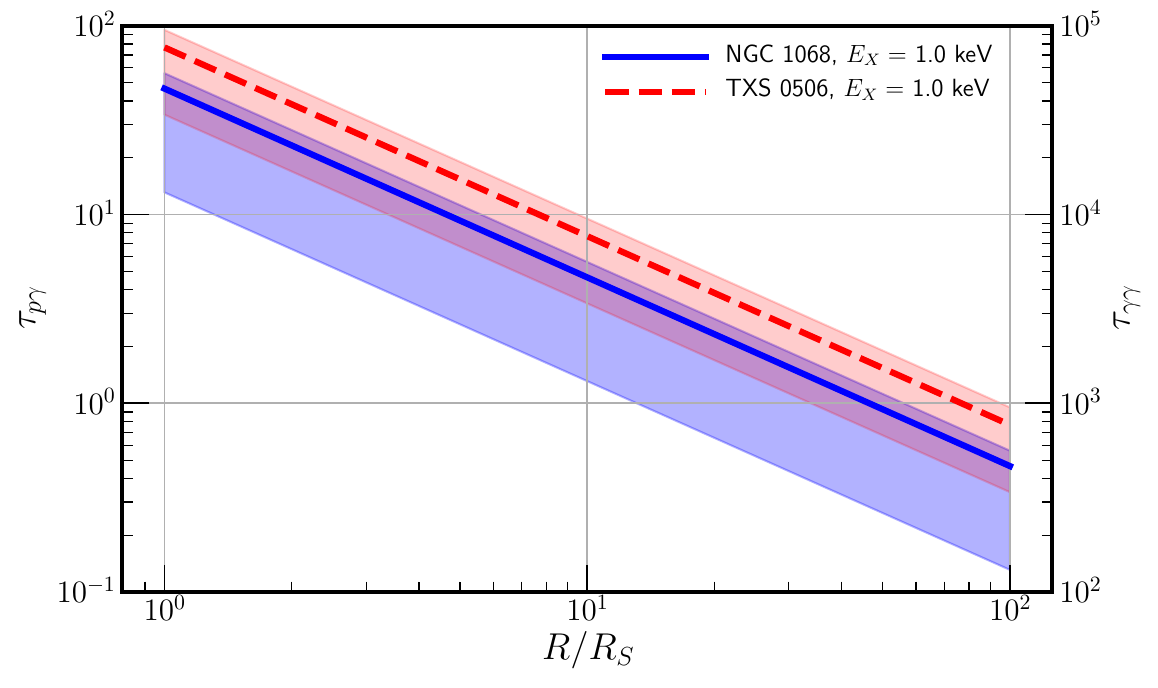}
\caption{Photopion opacities against emission radii for NGC~1068 (blue solid line) and TXS 0506 (red dashed line) computed using equation~\ref{eqn:tau_pg}. } 
\label{fig:tauvsR}
\end{figure}

NGC~1068 has a black hole mass of  $1.3 ^{+1.7}_{-0.6} \times 10^7$ M$_{\odot}$~\citep{ greenhill1997vlbi, kamenetzky2011dense, wang2020dynamical, padovani2024supermassive}, corresponding to an Eddington luminosity of $L_{\rm edd} = 1.6^{+2.1}_{-0.8} \times 10^{45}\,\rm erg\,s^{-1}$.  The intrinsic X-ray luminosity is reported to be $(3-7)\times 10^{43}\,\rm erg\,s^{-1}$ \citep{marinucci2015nustar, bauer2015nustar, padovani2024supermassive} in the $2-10$~keV band.
For a consistent comparison with the X-ray luminosity of TXS~0506+056, we extrapolate the $2-10$~keV flux to a broader $0.3-10$~keV X-ray band.  
By normalizing the intrinsic X-ray flux $F^{\text{intr}}_{X, \, 2-10 \, \rm keV} = 2.7 \times 10^{-10}\, \rm erg\,cm^{-2}\,s^{-1}$ from the latest BASS catalogue~\citep{ricci2017bat} and assuming a power-law spectral index of $\gamma_X=2.4^{+0.1}_{-0.1}$ based on the torus model, we obtain an intrinsic X-ray flux and luminosity of $F^{\text{intr}}_{X, \, 0.3-10 \, \rm keV} = 8.8^{+0.1}_{-0.8} \times 10^{-10}\, \rm erg\,cm^{-2}\,s^{-1}$. This corresponds to $L^{\text{intr}}_{X, \, 0.3-10 \, \rm keV} = 2.2^{+0.4}_{-1.7} \times 10^{43}  \,\rm erg\,s^{-1}$ for a source distance of $14.4$ Mpc \citep{bland1997ringberg, collaboration2019image, bottinelli1986malmquist}. 

The resulting opacity for $p\gamma$ interactions is shown in Fig.~\ref{fig:tauvsR}. The value of $\tau_{p\gamma} \sim 1$ for $R / R_S \sim 10$ reproduces the value obtained by more detailed modeling in the context of the core corona model~\citep{Murase:2022dog}. This dimensional analysis of NGC~1068 underscores the fact that the production of the neutrinos inside a dense X-ray core naturally reaches the level of neutrino production observed by IceCube. With $\tau_{\gamma\gamma} \sim 10^3 \times \tau_{p\gamma}$, as shown in Fig.~\ref{fig:tauvsR}, the gamma rays from the decay of neutral pions accompanying the neutrinos lose energy in the corona by pair production with the X-ray photons. The suppression accommodates the upper limits established by MAGIC~\citep{acciari2019constraints},
which limits their TeV gamma-ray flux to two orders of magnitude below the neutrino flux at some energies.    

The core of NGC~1068 not only hosts a dense X-ray corona but also dense clouds of hydrogen, identified by large values of $N_H$ along the line of sight~\citep{Rosas:2021zbx,Garcia-Burillo_2016}. The proton opacity due to interaction with the gas depends on the composition of the corona. If the proton energy density is comparable to or larger than the electron energy density in the corona, and if $\tau_{pp} \sim 1$, then neutrinos may actually be predominantly produced by pp interactions. Even if this is the case, the large value of $\tau_{p\gamma}$ is still required to absorb the gamma rays produced by the $\pi^0$ photons accompanying the observed neutrino flux.

Applying the same dimensional analysis to TXS~0506+056 results in similar opacity values as shown in Fig.~\ref{fig:tauvsR}. The literature agrees on a black hole mass of approximately $3 \times 10^8$ M$_{\odot}$~\citep{Padovani_2019, tjus2022neutrino}. In this work, we adopt the approach described in \citet{Padovani_2019} to infer the black hole mass with uncertainties. Taking the absolute R-band magnitude of the bulge to be $-22.9$, and using the active galactic black hole mass-bulge luminosity relation from \citep{mclure2002black}, we obtain a mass of $3.1^{+29.9}_{-2.7} \times 10^8$ M$_{\odot}$. This result is consistent with the previous estimate, and it corresponds to an Eddington luminosity of $L_{\rm edd} = 3.9^{+29.9}_{-3.4} \times 10^{46}\,\rm erg\,s^{-1}$.  To derive the X-ray luminosity, we adopt the {\it Swift}-XRT flux observation
of $F_{X, \, 0.3-10 \, \rm keV} = 2.3^{+0.5}_{-1.3} \times 10^{-12} \, \rm erg\,cm^{-2}\,s^{-1}$
in the $0.3-10$ keV band, which represents the mean flux when the source was in its flaring state \citep{keivani2018multimessenger}. Note that the observed X-ray flux of TXS 0506+056 serves as a lower limit to the intrinsic X-ray flux of the source, which includes contributions from the coronal, the jet, and other possible X-ray emitting components.
The intrinsic flux of the coronal region alone is rather unknown below 10~keV.
Assuming that it is comparable to this observed flux, we obtain an X-ray luminosity of $L^{\rm intr}_{X, \, 0.3-10 \, \rm keV} = 8.5^{+2.0}_{-4.8} \times 10^{44}\, \rm erg \,s^{-1}$ assuming a source luminosity distance of $d_L = 1774$ Mpc \citep{paiano2018redshift}. As shown in Fig.~\ref{fig:tauvsR}, a value of $\tau_{p\gamma} \sim 1$ is also found for $R/R_S \sim 10$, which decreases to 0.1 for $R/R_S \sim 100$.

Next, we compute the proton emissivity $Q_p$ required to generate the observed neutrino emissivities, $Q_\nu$, in a target with opacity $\tau_{p\gamma}$:

\begin{equation}
    E_\nu^2  Q_\nu \approx \frac{3}{8} f_{p\gamma} E_p^2 Q_p\,,
    \label{CRlum1}
\end{equation}
where  
\begin{equation}
    f_{p\gamma} \equiv 1 - e^{-\tau_{p\gamma}} \,.
    \label{CRlum2}
\end{equation}
is the pion production efficiency. In this equation, the differential neutrino luminosity $L_\nu = E_\nu^2 Q_\nu$ includes the three neutrino flavors.

\begin{figure}[t]
\centering
\includegraphics[width=0.48\textwidth]{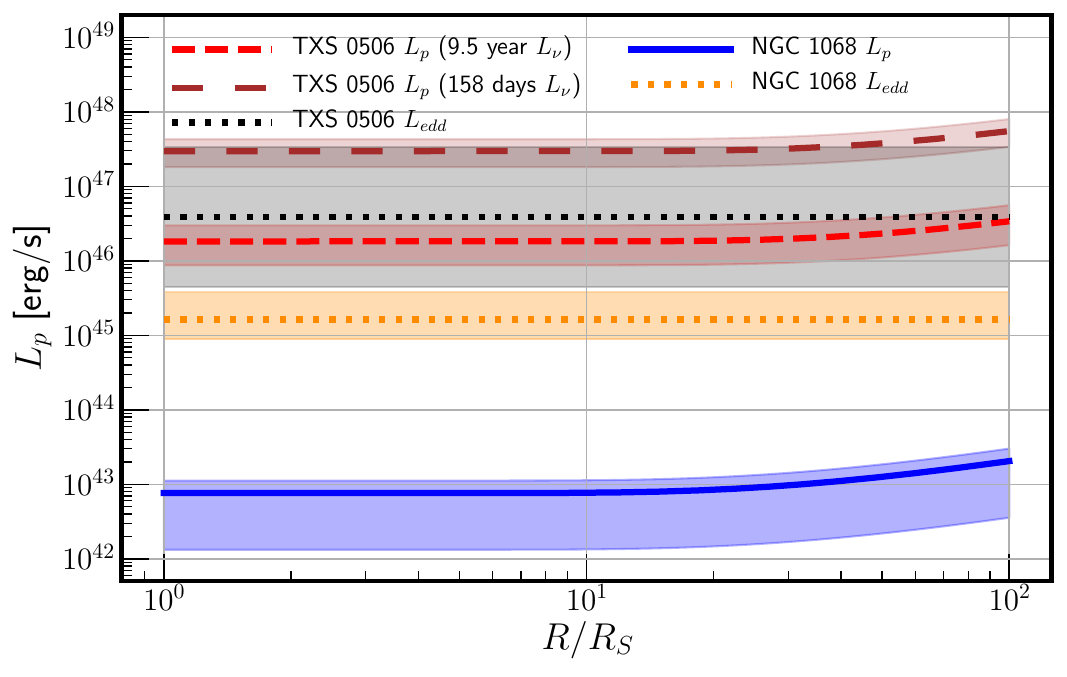}
\caption{Plot of proton and Eddington luminosities against emission radii for NGC~1068 (blue solid and orange dotted lines) and TXS~0506+056 (red dashed, brown loosely dashed and black dotted lines). The red dashed and brown loosely dashed lines denote the proton luminosities computed using 9.5 years and 158 days of neutrino data respectively.} 
\label{fig:LpvsR}
\end{figure}

At large neutrino emission radii $R\gg R_s$ and small opacity limit $\tau_{p\gamma} \ll 1$, Eq.~\ref{CRlum1} yields
\begin{align}
    &L_\nu 
    \approx \frac{\kappa_{p\gamma}\sigma_{p\gamma}}{4\pi v_{\rm esc}}\frac{L_p}{R E_X}L_X 
    \nonumber \\
    &= 0.7 \left(\frac{v_{\rm esc}}
    {0.1\,c}\right)^{-1}\left(\frac{L_p}{10^{-2}\,L_{\rm edd}}\right)\left(\frac{R}{10\,R_s}\right)^{-1}\left(\frac{E_X}{1\,\rm keV}\right)^{-1} L_X .
    \label{Eq: LnuLX}
\end{align}

\begin{figure}[h!]
\centering
\includegraphics[width=0.48\textwidth]{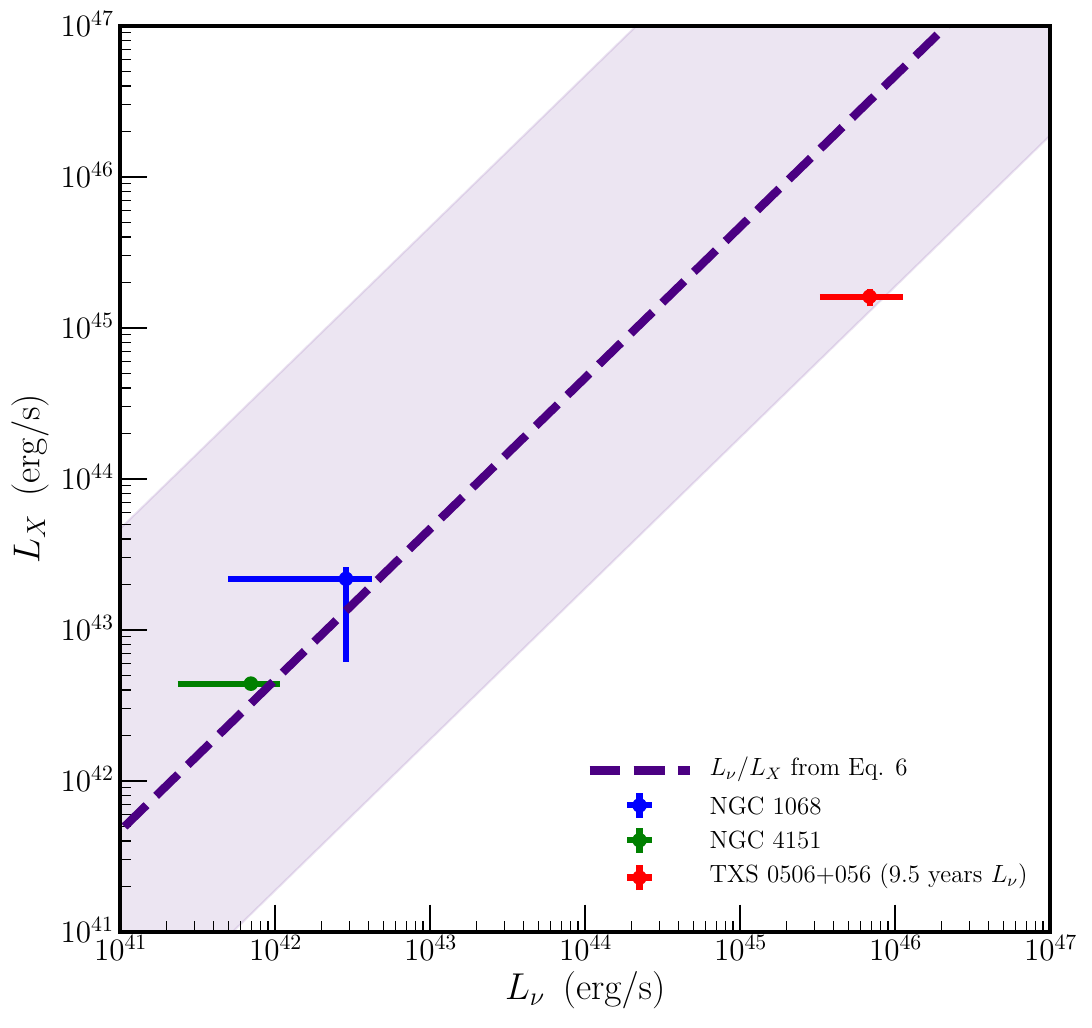}
\caption{
Plot showing the X-ray and neutrino luminosities for sources in steady emission states, namely NGC~1068 (blue), TXS~0506+056 (red) and NGC~4151 (green). The red point is plotted using the neutrino luminosity computed using the 9.5 years of IceCube data and the NuSTAR X-ray fluxes reported in \citep{keivani2018multimessenger}. The purple line shows the linear relationship between $L_\nu$ and $L_X$ derived in Eq.~(\ref{Eq: LnuLX}), and the purple shaded region encodes the propagated errors associated with the slope, where we took the ranges of input parameters $L_p/L_{\rm edd}$ to be between $10^{-3}$ and $1$, and $R/R_S$ between $10-100$. 
}
\label{fig: LinearLxLnu}
\end{figure}

In Figure~\ref{fig: LinearLxLnu}, we compare Eq.~(\ref{Eq: LnuLX}) with observations of sources assuming that they produce steady neutrino emission. For NGC~4151 \citep{inoue2023gamma, yuan2020cepheid, bentz2015agn, bentz2022broad}, we calculated the neutrino luminosity using the flux and spectral index reported in \citet{abbasi2024searchneutrinoemissionhard} while the X-ray flux is determined using the same methodology as for NGC~1068, with parameters obtained from \citet{ricci2017bat}. Figure~\ref{fig: LinearLxLnu} indicates that the previously discussed analysis would also apply to the case of NGC~4151.
References~\citet{kun2024correlation, neronov2024neutrino} have also argued for a neutrino luminosity that is proportional to the X-ray luminosity with a coefficient of order unity. 

In the large opacity limit $\tau_{p\gamma} \gg 1$, the exponential term gets suppressed, and we are left with a neutrino luminosity that scales linearly with proton luminosity,
\begin{align}
    L_\nu \sim \frac{3}{8} L_p \,.
\end{align}

For NGC~1068, the all-flavor neutrino flux is $\phi_0 = (15.0 \pm 6.3) \times 10^{-11}$ TeV$^{-1}$ cm$^{-2}$ s$^{-1}$ at 1~TeV with a spectral index of $\gamma = 3.2 \pm 0.2$~\citep{icecube2022evidence}. Integrating the flux between neutrino energies of $1.5$ and 15 TeV~\citep{icecube2022evidence} yields a neutrino luminosity of $L_\nu = 2.9^{+1.4}_{-2.4} \times 10^{42}\,\rm erg\,s^{-1}$. To produce such a neutrino flux, a proton luminosity of 
$L_p = 7.7^{+3.6}_{-6.4} \times 10^{42}\,\rm erg\,s^{-1}$ 
is required, assuming an neutrino emission radius of $R/R_S = 10^{+90}_{-9}$. 

For TXS~0506$+$056, a 9.5-year time-integrated neutrino flux of $\Phi_{\nu_{\mu}} = 0.8^{+0.5}_{-0.4} \times 10^{-16}$ TeV$^{-1}$ cm$^{-2}$ s$^{-1}$ normalized at $100$~TeV, and a spectral index of $\gamma = 2.0 \pm 0.3$ at 100~TeV~\citep{icecube2018multimessenger} corresponds to a neutrino luminosity of $L_{\nu_\mu} = 2.3^{+1.5}_{-1.2} \times 10^{45} \ \rm erg\,s^{-1}$ between 32 TeV and 3.6 PeV. At the same neutrino emission radius of $R/R_S = 10^{+90}_{-9}$,  a proton luminosity of 
$L_p = 1.8^{+1.2}_{-1.0} \times 10^{46} \ \rm erg\,s^{-1}$.
is needed assuming that the neutrino emission is isotropic over a $4\pi$ solid angle. Alternatively, we could use the 158 days neutrino flux (observed in 2014) of $\Phi_{\nu_\mu} = 1.6^{+0.7}_{-0.6} \times 10^{-15}$ TeV$^{-1}$ cm$^{-2}$ s$^{-1}$ normalized at $100$~TeV with a spectral index of $\gamma = 2.2\pm0.2$~\citep{icecube2018multimessenger}. This corresponds to a neutrino luminosity of $L_{\nu_\mu} = 3.7^{+1.7}_{-1.4} \times 10^{46} \ \rm erg\,s^{-1}$ when integrated between 32 TeV and 3.6 PeV and a proton luminosity of $L_p = 3.0^{+1.3}_{-1.1} \times 10^{47} \ \rm erg\,s^{-1}$.

The proton luminosities for emission radii between $R/R_S = 1 - 100$ are shown in Fig.~\ref{fig:LpvsR} for NGC~1068 (blue solid line) and TXS~0506+056 (red dashed and brown loosely dashed lines). Note that while the proton luminosity is well below Eddington limit for NGC~1068, to explain the time-integrated flux of TXS~0506+056, $L_p \sim 20 \,L_X$ is needed. To explain the flares of TXS~0506+056, it requires the presence of a corona with Eddington-level power.  This is a challenging condition to meet, requiring scenarios where the corona power is dominated by relativistic protons and a super-Eddington activity that occurred around the time of IC-170922. Alternatively, if one constrains the proton luminosity by $L_p \lesssim L_X$, our model predicts that the corona of TXS~0506+056 may power a neutrino luminosity up to $L_\nu \sim 3.4\times 10^{44}\,\rm erg\,s^{-1}$.


\section{Conclusion}
\label{sec:concl}

The production of high-energy neutrinos in the dense cores of active galaxies could potentially provide a unified explanation for the production mechanism in the first two neutrino sources observed by IceCube. Although qualitative, our analysis meets the challenge of modeling the TXS~0506+056 multimessenger spectrum. A more detailed analysis of the data is made complicated by the fact that for production of neutrinos near the dense cores of active galaxies, it is likely that neutrinos are not only produced in the X-ray corona but also on the high hydrogen density near the core as well as on UV photons associated with the accretion disk. As already mentioned for NGC~1068, the measurement of the optical depth, $N_H$, implies that $\tau_{pp} > \tau_{p\gamma}$ and both mechanisms contribute to the observed flux. Importantly, even in the presence of a significant $pp$ contribution, the presence of the corona is essential for absorbing the pionic gamma rays, which are not observed at TeV energies. Also, our modeling for TXS~0506+056 is concentrated on the 2014 burst, which produced the bulk of the neutrino flux over the time that IceCube had taken data. The subdominant hour-long 2017 burst is dominated by a single high-energy neutrino, which is more naturally produced on a UV photon than on an X-ray. In any case, our proposal emphasizes the importance of the X-ray flux in selecting potential neutrino sources, a tool that has already been successfully explored by IceCube to find evidence for more sources~\citep{abbasi2024icecubesearchneutrinoemission,abbasi2024searchneutrinoemissionhard}.

Finally, we entertain the possibility that the sources discussed above represent generic sources where extragalactic cosmic rays originate. The relation~\citep{Ahlers:2014ioa} connecting the flux of the nearest source with neutrino flux $L_\nu$ to the total diffuse flux $\Phi_\nu$ from a uniform distribution of sources with a density $\rho(z)$ in the Universe is given by
\begin{equation}
{E_\nu^2}\Phi_\nu^{\text{diff}} = \frac{1}{4\pi} \frac{c}{H_0} \xi_z \, \rho L_\nu\,,
\end{equation}
where $H_0$ is the Hubble parameter and $\xi_z$ the result of integration over the redshift of the sources, which reduces to an overall factor assuming that the fluxes follow a power law; for instance, $\xi_z \simeq 0.5 $ for no source evolution or $\xi_z \simeq 2.6 $ for star formation evolution \citep{yuksel2008revealing}. For the value of the diffuse astrophysical neutrino flux observed by IceCube with $\Phi^{\rm Astro}_\nu = 5.04 \times 10^{-11} \rm TeV^{-1} cm^{-2} s^{-1} sr^{-1}$ at 100~TeV \citep{abbasi2024characterization}, we find that the local source density of NGC~1068-like and TXS~0506+056-like sources is $\rho_0 = 1.0 \times 10^{-6}$ Mpc$^{-3}$ and $\rho_0 = 1.5  \times 10^{-9}$ Mpc$^{-3}$ ($\rho_0 = 2.0  \times 10^{-7}$ Mpc$^{-3}$ and $\rho_0 = 2.8  \times 10^{-10}$ Mpc$^{-3}$ ),
respectively, assuming star formation evolution (no source evolution).  With densities $\rho_0 \sim 10^{-6}$ Mpc$^{-3}$ for NGC~1068-like sources, the diffuse flux is accommodated by $\sim 10^3$ sources, which also happens to be the number of sources in the Universe with X-ray luminosity in excess of $10^{43}$ erg s$^{-1}$~\citep{Urry_1995}. For an episodic source like TXS~0506+056, the fraction of the time is not known for which it contributes at the high level of the Eddington-level 100-day 2014 burst, which dominates the total emission in the decade that IceCube has observed the source. With an on-time of $5\%$, the density of the sources matches that of NGC~1068~\citep{halzen2019neutrino}.

In this context, it is also intriguing that with $\tau_{p\gamma}$ values of order unity for both sources, the diffuse neutrino flux matches the total energy loss $\rho_pL_p \sim 10^{44}\,\rm erg \, Mpc^{-3} \, yr^{-1}$ of ultrahigh-energy cosmic rays in the Universe~\citep{katz2013energy} measured by the Auger experiment and Telescope Array. 

\begin{acknowledgments}
This work is supported by the Office of the Vice Chancellor for Research at the University of Wisconsin--Madison with funding from the Wisconsin Alumni Research Foundation. K.F. acknowledges support from the National Science Foundation (PHY-2238916). This work was supported by a grant from the Simons Foundation (00001470, KF). K.F acknowledges the support of the Sloan Research Fellowship. The research of F.H was also supported in part by the U.S. National Science Foundation under grants~PHY-2209445 and OPP-2042807 and by the Balzan Foundation.
\end{acknowledgments}


\bibliography{biblio}{}
\bibliographystyle{aasjournal}

\end{document}